\documentclass[fleqn,usenatbib]{mnras}

\usepackage{graphicx}
\usepackage{times}
\usepackage{subfig}
\usepackage{indentfirst}
\usepackage{amssymb} 
\usepackage{txfonts}
\usepackage{appendix}
\usepackage{tabu,multirow}
\usepackage{textcomp}
\usepackage{natbib}
\usepackage{longtable, portland}
\usepackage{supertabular}
\usepackage{rotating}
\usepackage{array}
\usepackage{caption}
\usepackage{blindtext}
\usepackage{gensymb}
\usepackage[T1]{fontenc}
\usepackage{color}
\usepackage{cleveref}
\defcitealias{Mahoro2019}{M19}
\newcommand\angstrom{\mbox{\normalfont\AA}}

\title[Green valley active and inactive galaxies]{Stellar populations of a sample of far-IR AGN and non-AGN green valley galaxies}

\author[A. Mahoro et al.]{Antoine Mahoro $^{1,\,2}$\thanks{E-mail: antoine@saao.ac.za}, Mirjana Povi\'c$^{3,\,4,\,5}$, Petri V\"ais\"anen$^{1,\,6}$, Pheneas Nkundabakura$^{7}$, and \newauthor Kurt van der Heyden$^{2,\,8}$\\
$^{1}$South African Astronomical Observatory, P.O. Box 9 Observatory, Cape Town, South Africa\\
$^{2}$Department of Astronomy, University of Cape Town, Private Bag X3, Rondebosch 7701, South Africa\\
$^{3}$Astronomy and Astrophysics Research and Development Department, Entoto Observatory and Research Center (EORC), \\
 Ethiopian Space Science and Technology Institute (ESSTI), P.O. Box 33679, Addis Ababa, Ethiopia\\
$^{4}$Instituto de Astrof\'isica de Andaluc\'ia (IAA-CSIC), Glorieta de la Astronom\'ia s/n, 18008 Granada, Spain\\
$^{5}$Physics Department, Faculty of Science, Mbarara University of Science and Technology, P.O. BOX: 1410 Mbarara, Uganda\\
$^{6}$Southern African Large Telescope, P.O. Box 9 Observatory, Cape Town, South Africa\\
$^{7}$MSPE Department, School of Education, College of Education, University of Rwanda,  P.O. Box 5039, Kigali, Rwanda\\
$^{8}$National Research Foundation, Box 2600, Pretoria, 0001, South Africa}
\date{Accepted 2022 April 22. Received 2022 April 8; in original form 2021 August 16}

\pubyear{2021}

\begin{document}
\label{firstpage}
\pagerange{\pageref{firstpage}--\pageref{lastpage}}
\maketitle
\begin{abstract}
We present a study on the stellar populations and stellar ages of a sub-sample of far-infrared AGN and non-AGN green valley galaxies analysed in \cite{Mahoro2017, Mahoro2019} at $\rm {0.6\,<\,z\,<\,1.0}$ using the data from the COSMOS field. We used long-slit spectroscopy and derived stellar populations and stellar ages using the stellar population synthesis code “STARLIGHT” and analysed the available Lick/IDS indices, such as Dn4000 and H${\delta}_{A}$. We find that both FIR AGN and non-AGN green valley galaxies are dominated by intermediate stellar populations 67\% and 53\%, respectively. The median stellar ages for AGN and non-AGN are $\rm{\log t\,=\,8.5\,[yr]}$ and $\rm{\log t\,=\,8.4\,[yr]}$, respectively. We found that majority of our sources (62\% of AGN and 66\% of non-AGN) could have experienced bursts and continuous star formation. In addition, most of our FIR AGN (38\%) compared to FIR non-AGN (27\%) might have experienced a burst of SF more than 0.1\,Gyr ago. We also found that our FIR AGN and non-AGN green valley galaxies have similar quenching time-scales of $\sim$\,70\,Myr. Therefore, the results obtained here are in line with our previous results where we do not find that our sample of FIR AGN in the green valley shows signs of negative AGN feedback, as has been suggested previously in optical studies.  

\end{abstract}




\section{Introduction}\label{Into}
Studies of a large sample of galaxies at different redshifts have demonstrated the existence of a bi-modality in terms of different properties, such as morphology, colours, stellar mass, star formation rates (SFR), etc. \citep[e.g.,][]{Kauffmann2003, Baldry2004, Salim2007, Brammer2009, Mendez2011, Povic2013, Walker2013, Lee2015, Bremer2018, Angthopo2019, NogueiraCavalcante2019, Nyiransengiyumva2020, Jian2020, Das2021}. This bi-modality differentiates between two dominant regions: the “blue cloud”, populated mainly by late-type galaxies (LTGs), commonly characterised by blue colours and spiral or irregular morphologies, and the “red sequence”, populated mainly by early-type galaxies (ETGs) with red colours and elliptical or lenticular morphologies.

In between the blue cloud and red sequence, one encounters a less densely populated region, commonly referred to as the “green valley,” which is thought to represent transitional stage in galaxy evolution from LTGs to ETGs \citep[e.g.,][and references therein]{Salim2014}. The physical properties of galaxies in the green valley region differ from those in either the blue cloud or the red sequence, indicating intermediate characteristics including their morphological parameters \citep[e.g.,][]{Mendez2011, Pan2013, Gu2018, Gu2019, Ge2019}, luminosity and stellar mass \citep[e.g.,][]{Goncalves2012, Phillipps2019}, age of stellar populations \citep[e.g.,][]{Pan2013, Phillipps2019, Angthopo2020}, gas properties \citep[e.g.,][]{Schawinski2014}, or metallicity \citep[e.g.,][]{Angthopo2020}. Therefore, understanding the properties of green valley galaxies is essential to a better understanding of how galaxies evolve from LTGs to ETGs. However, there are also still many inconsistencies found in different studies regarding properties of green valley galaxies, such as their environments \citep[e.g.,][]{Pan2013, Salim2014, Lee2015, Coenda2018, Ge2019, Jian2020, Das2021}, fraction \citep[e.g.,][]{Schawinski2014, Jian2020, Das2021}, or structure \citep[e.g.,][]{Schawinski2014, Das2021}. 

Stellar population analysis aims to understand a galaxy’s formation and evolution by studying properties such as stellar types, kinematics, chemical abundances, ages, and even to infer metallicities of its constituent luminous parts. The analysis of metallicity in galaxies is, however, complicated by the fact that observables such as broadband colours and spectral line indices, which are sensitive to chemical enrichment, also respond in a similar way to changes in luminosity-weighted age. This is famously known as the age-metallicity degeneracy \citep[e.g.,][]{Worthey1994,Bruzual2003,Spengler2017,Liu2020}. 
Stellar population studies through the spectra of galaxies can be carried out using an approach called stellar population synthesis \citep[e.g.,][]{Bruzual2003, Vazdekis2010, DazGarca2015, MenesesGoytia2015, Byler2017, CidFernandes2018}. Several authors have developed various stellar population synthesis models using the evolutionary population synthesis approach. Various parameters, such as initial stellar mass function, chemical composition, star formation histories, and metallicity can be measured \citep[e.g.,][]{Bruzual2003, Conroy2010, Conroy2012, Conroy2013, Fioc1997, Maraston1998, Maraston2005, Vazdekis1996, Vazdekis2015, Fioc2019, Hosek2020}.

Previous works studied stellar population properties of ETGs. These identified the dominance of the older stellar populations \citep[e.g.][]{Lotz2000, Renzini2006, Gargiulo2012, Silchenko2012, Dahmer-Hahn2018, FerreMateu2019, Lonoce2020, Zibetti2020, DahmerHahn2022}, which had a total metallicity between less than half-solar to higher values \citep[e.g.][]{Lotz2000, Lonoce2020, Zibetti2020} and enhanced $\alpha$-elements over iron \citep[e.g.][]{Colucci2013, Parikh2019}. These results are in line with \cite{Goddard2017} who used the SDSS-IV MaNGA integral field spectroscopic data and analysed stellar population parameters of ETGs as a function of radius. They found that ETGs generally have older stellar populations with a slight light-weighted age gradient.Similar was obtained in the CALIFA (Calar Alto Legacy Integral Field Area) survey, finding that ETGs have negative luminosity weighted mean age gradients with radius \citep{GonzlezDelgado2015, Zibetti2017}.

Many studies on stellar population properties were performed using star-forming galaxies and starburst galaxies from optical data. These conclude that H II galaxies are best described by episodic stellar populations, including old, intermediate, and younger populations \citep[e.g.][]{ Bonatto2000, CidFernandes2003, Chen2009, Martins2013, Dametto2014,  Telles2018, Weistrop2020}. Using near-infrared spectroscopy, \cite{Riffel2008} analysed central (inner few hundred parsecs) stellar populations of starburst galaxies and found that the significant population was a 1 Gyr old with solar metallicity component.
Previous works studied stellar populations of green valley galaxies finding on average older populations than those in blue cloud and younger than those in red sequence galaxies \citep[e.g.,][]{Pan2013, Angthopo2020}. \cite{Trussler2020} analysed mass-weighted stellar age and metallicity of green valley and found that green valley galaxies have intermediate properties between star-forming and passive galaxies in both metallicity and age.

Furthermore, studies done on active galactic nucleus (AGN) sources found that in AGN hosts with more luminous AGN the contribution of younger stellar populations to the optical emission is larger than that for low-luminosity AGN \citep[e.g.,][]{Rembold2017}. Also, some works focused on the stellar populations of quasar's host galaxies and found that they contain younger stellar populations \citep[e.g.,][]{Canalizo2013, Sanmartim2013, Mosby2015, Bessiere2017}. Studies of the stellar populations of low-luminosity active galactic nuclei (LLAGN) found that their young stars contribute very little to the optical continuum \citep[e.g.,][]{Kauffmann2003b, GonzlezDelgado2004, CidFernandes2004, CidFernandes2005, Povic2016}. In contrast, intermediate-age stars contribute significantly, and most of the strong [OI] LLAGNs have predominantly older stellar populations.

On the other hand, NIR studies of AGN hosts found the signs of stellar rings with ages 0.1\,-\,2.0\,Gyr \citep[e.g.][]{Riffel2010, Riffel2011, DahmerHahn2019, Diniz2019,Riffel2022}. The authors also found that these regions are
generally related to rings with lower stellar dispersions, proposing that the intermediate-age stellar populations were formed from low-velocity dispersion gas.

Recently, \cite{Mallmann2018} studied the stellar population properties of AGN hosts at $\rm{\langle\,z\,\rangle}\,\rm{\approx}$ 0.03 and compared the result with a control sample of non-active galaxies. They found that the fraction of young stellar populations is higher in the inner regions ($\rm{R\,\leq\,0.5R_{e}}$) of high-luminosity AGN than in the case of non-active galaxies. In addition, low-luminosity AGN present similar younger star fractions to the control sample hosts, for the entire studied range ($1R_{e}$). The fraction of the AGN hosts' intermediate-age stellar populations increases outwards, with an apparent enhancement compared with the control sample. They also found that galaxies' inner regions (AGN and control galaxies) present predominantly older stellar populations, whose fraction decreases outwards. 

Interestingly, the rate of AGN detection is high in green valley galaxies \citep[e.g.,][]{Nandra2007, Schawinski2010, Povic2012, Wang2017, Gu2018, Lacerda2020}, suggesting a link between AGN activity and star-formation quenching, moving galaxies from the blue cloud to the red sequence. For this reason, it is important to compare the properties of active and non-active galaxies in the green valley to understand fully their nature and the role of AGN in galaxy evolution. In line with this, we studied a sample of green valley active and non-active galaxies with infrared emission selected in the Cosmological Evolution Survey (COSMOS\footnote{http://cosmos.astro.caltech.edu/}) \citep{Scoville2007}. As reported in \cite{Mahoro2017}, we measured and analysed the SFR properties of both far-infrared (FIR) AGN and FIR non-AGN green valley galaxies. We observed that the green valley AGN with FIR emission do not show signs of SF quenching since 68\% and 14\% of sources are located on and above the main sequence (MS) of SF, respectively. We found that 70\% of non-AGN FIR green valley galaxies are on the MS, with 9\% being above it. Within the same stellar mass range, we found that AGN galaxies in our sample have enhanced SFRs in comparison to non-AGN, independently of their morphology \citep{Mahoro2017}, suggesting that if there is any influence of AGN on star-formation properties of their host galaxies we observe signs of positive AGN feedback rather than the negative one which has been suggested in most of X-ray and optical studies \citep[e.g.,][]{Nandra2007,Povic2012,Shimizu2015,Leslie2016,Lin2019}.

In this work, we want to go a step further and better understand the properties of stellar populations and ages of our sample of FIR AGN and non-AGN using optical spectroscopic data. This will help us better understand previously obtained results, but also to remove any discrepancy that previously has been found when measuring SFRs in optical and FIR \citep[e.g.,][]{Lee2013, Hayward2014, Davies2016, Ellison2016, Katsianis2017}. We use data from the Large Early Galaxy Astrophysics Census (LEGA-C) survey \citep{vanderWel2016} to study the stellar population of green valley galaxies through the  stellar population synthesis. 

The paper is organised as follows: Data are presented in Section 2, while methodology is explained in Section 3. In Section 4 we present our main analysis and results, and Section 5 discusses the obtained results. Section 6 gives the summary of the paper.

Throughout the paper, we assume an $\Omega_{m}=0.3,\,\Omega_{\Lambda}=0.7$, with $H_{0}=70$\,km\,s$^{-1}$\,Mpc$ ^{-1}$ cosmology. All magnitudes are in the AB system. The stellar masses are given in units of solar masses (M$_{\odot}$), and both SFR and stellar masses assume \cite{Salpeter1955} initial mass function (IMF).

\section{Data}\label{data}

This study continues the work presented in \cite{Mahoro2017} and \cite{Mahoro2019} based on a sample of photometric data taken in the COSMOS field \citep{Scoville2007}. The initial sample is selected from \cite{Tasca2009}, with a magnitude completeness of I\, =\,23. 

We selected our AGN sample using two major X-ray  \textit{XMM-Newton} and \textit{Chandra} public catalogues \citep{Brusa2007, Civano2012}. The ratio between the X-ray flux in the hard $\rm{2\,-\,10\,keV}$ band and the optical I band flux within the range of $\rm{-1\leq\,logF_{x}/F_{O}\leq1}$ was applied to select AGN \citep[e.g.,][]{Alexander2001, Bauer2004, Trump2009}. 
Our green valley galaxies sample was selected using $\rm{U-B}$ rest-frame colour with the criterion of $\rm{0.8\leq U-B\leq 1.2}$ \citep[e.g.][]{Willmer2006, Nandra2007,Bornancini2018}.

The infrared (IR) luminosities and SFRs were obtained using \textit{Herschel}/PACS 160, 100 $\micron$ \citep{Lutz2011} and \textit{Spitzer} DR1 24 $\micron$ \citep{Rieke2004} flux densities through spectral energy distribution (SED) fitting by using the Le Phare\footnote{http://www.cfht.hawaii.edu/~arnouts/LE PHARE/lephare.html.} code \citep{Arnouts2011, Ilbert2006}. The fitting was made using two different templates. AGN sources were fitted using \cite{Kirkpatrick2015} templates where we were able to correct IR luminosities for AGN contribution \citep{Mahoro2017}. Non-AGN sources were fitted using \cite{Chary2001} templates. 

After checking all LePhare output fits visually, we obtained a final sample of 103 AGN and 2609 non-AGN green valley galaxies for our analysis. In this work, this sample we will call the initial sample \citepalias[hereafter][]{Mahoro2019}. The used stellar masses were measured through SED fitting using the information from 10 optical/NIR photometric bands and the KCORRECT code \citep{Blanton2007}.

\subsection{Optical spectroscopic data}\label{specdata}

To study the spectroscopic properties of AGN and non-AGN green valley galaxies described above, we cross-matched the \citetalias{Mahoro2019} sample with data from the LEGA-C survey \citep{vanderWel2016}. LEGA-C is a  VLT/VIMOS European Southern Observatory (ESO) public spectroscopic survey of $\sim$ 3000 galaxies in the COSMOS field with a redshifts range of $\rm{0.6\,<\,z\,<\,1}$. Each galaxy was observed for $\sim$\,20 hours, which results in spectra with S/N $\sim$\,20 $\angstrom ^{-1}$ (with resolution R $\sim$\,3000) in the wavelength range $\sim\, 0.6\,\micron\,-\,0.9\,\micron$.  In this work, we use the second data release (DR2)\footnote{https://www2.mpia-hd.mpg.de/home/legac/}, consisting of 1988 spectra in total \citep{Straatman2018}. A total of 17 AGN and 191 non-AGN counterparts were found when our  \citetalias{Mahoro2019} sample was cross-matched with the \cite{Straatman2018} catalogue. This sample is the final sample of analysis, and it is called M20 in this paper. We note that the M20 sample obtained is in the mass range of $\rm{\log\,M_{*}\,=\,10.6M_{\odot}-11.6M_{\odot}}$. 
\subsubsection{Comparison between M20 and \citetalias{Mahoro2019} samples}\label{M19_M20_comparsion}

This section presents comparison of physical properties of both M20 and \citetalias{Mahoro2019} samples, for better understanding of our M20 subsample. \autoref{fig_parameters_comp} shows the comparison of five physical parameters: morphological type, photometric redshift, SFR, stellar mass, and F814W magnitude of the M20 subsample and \citetalias{Mahoro2019} sample (left and middle plots). \autoref{TableM19vsM20comp} shows the median values, first quartile (Q1, covering 25\% of samples) and third quartile (Q3, covering 75\% of samples) of all analysed parameters measured for the two samples. 

Using morphological classification from \cite{Mahoro2019}, we compared the morphological types of M20 and \citetalias{Mahoro2019} samples, as can be seen in the top plot of \autoref{fig_parameters_comp}. Morphological types are: class 1 for elliptical, S0 or S0/S0a galaxies, class 2 for spirals, class 3 for irregulars, class 4 for peculiar galaxies and finally class 5 indicates unclassified galaxies \citep{Mahoro2019}. In terms of morphology, we do not find any significant difference in the M20 subsample in comparison to the \citetalias{Mahoro2019} sample.

However, as can be seen in \autoref{fig_parameters_comp}, the two samples show some differences in redshift, SFR, $\rm{M_{\odot}} $, and magnitude distributions.
In the case of redshift, the M20 sample is at slightly higher redshift, with 50\% of AGN (non-AGN) being between $\rm{0.73\,-\,0.93}$ $\rm{(0.67\,-\,0.88)}$, while the \citetalias{Mahoro2019} sample has values between $\rm{0.59\,-\,0.94\, (0.34\,-\,0.8)}$, respectively.
The stellar mass distribution shows that 50\% of AGN (non-AGN) have values between $\rm{11.11\,-\,11.29}$ $\rm{(10.9\,-\,11.23)}$ and  $\rm{10.88\,-\,11.27\,(10.3\,-\,10.96}$) in the M20 and \citetalias{Mahoro2019} samples, respectively. We also found that AGN and non-AGN in the M20 subsample have slightly higher SFRs than the \citetalias{Mahoro2019} sample, with 50\% of AGN (non-AGN) having values between $\rm{63.0\,-\,131.8\,(33.8\,-\,81.2)}$. Finally, we obtain that the M20 subsample is slightly fainter in comparison to the \citetalias{Mahoro2019} sample, where 50\% of AGN and non-AGN have magnitudes in the range $\rm{21.1\,-\,21.4}$ and $\rm{21.0\,-\,21.8}$, respectively. These small differences between the two samples are a natural consequence of correlating the \citetalias{Mahoro2019} sample with a K-band flux selected LEGA-C survey that was further restricted to a specific redshift range of $\rm{z\,=\,0.6\,-\,1.0}$.

\begin{table*}
\caption{Comparison of photometric redshift, SFR, stellar mass, and magnitude properties between the M20 and ~\protect\citetalias{Mahoro2019} samples of FIR AGN and non-AGN galaxies.}
\begin{tabular}{cccc|ccc|ccc|ccc}\\ \hline
                             & \multicolumn{6}{c|}{FIR AGNs}                           & \multicolumn{6}{c}{FIR Non-AGNs}                       \\ \hline
                             & \multicolumn{3}{c|}{M20}& \multicolumn{3}{c|}{\citetalias{Mahoro2019}} & \multicolumn{3}{c|}{M20} & \multicolumn{3}{c}{\citetalias{Mahoro2019}} \\ 
                             & Q1     & M      & Q3      & Q1      & M       & Q3      & Q1     & M       & Q3      & Q1     & M      & Q3    \\ \hline
z                            & 0.73    & 0.85    & 0.93  & 0.59    & 0.83    & 0.94    &  0.67  & 0.73   & 0.88    & 0.34   & 0.54   & 0.80     \\
$\rm{\log M_{*}[M_{\odot}]}$            & 11.11   & 11.16   & 11.29 & 10.88   & 11.11   & 11.27   &  10.90 & 11.09  & 11.23   & 10.03  & 10.66  & 10.96    \\
SFR $\rm{[M_{\odot}/yr]}$    & 63   & 87.09   & 131.82& 30.9    & 63   & 128.82  &  33.88 & 48.97  & 81.28   & 8.51   & 23.98  & 50.11     \\
mag(F814W)                   & 21.17   & 21.20   & 21.45 & 20.95   & 21.39   & 22.11   &  21.05 & 21.57  & 21.82   & 20.47  & 21.30  & 22.01    \\ \hline
\end{tabular}
\label{TableM19vsM20comp}
\end{table*}

\begin{table}
\centering
\caption{Number of galaxies (and fraction) in morphological visual classification from \protect\cite{Mahoro2019}.}
\label{tab_Visual_number}
\begin{tabular}{ccccc} \hline
        & \multicolumn{2}{c}{FIR AGNs} & \multicolumn{2}{c}{FIR non-AGNs}  \\
        & M20       & \citetalias{Mahoro2019}  & M20& \citetalias{Mahoro2019}\\ \hline
Class 1 & 3 (23\%)  & 26 (25 \%)               & 24 (22\%)  & 452 (17\%)            \\
Class 2 & 5 (38\% ) & 27 (26 \%)               & 50 (45\%)  & 1204 (46\%)          \\
Class 3 & -         & 2 (2\%)                  & -          & 87 (3\%)              \\
Class 4 & 3 (23\% ) & 39 (38 \%)               & 27 (25\%)  & 494 (19\%)            \\
Class 5 & 2 (15 \%) & 9 (9\% )                 & 9 (8\%)    & 372 (14\%)             \\ \hline
\end{tabular}
\end{table}

\subsubsection{Comparison of AGN and non-AGN in M20 sample.}\label{AGN_NON_AGN_m20_sample}

\begin{figure*}
\centering
\includegraphics[width=0.99\textwidth]{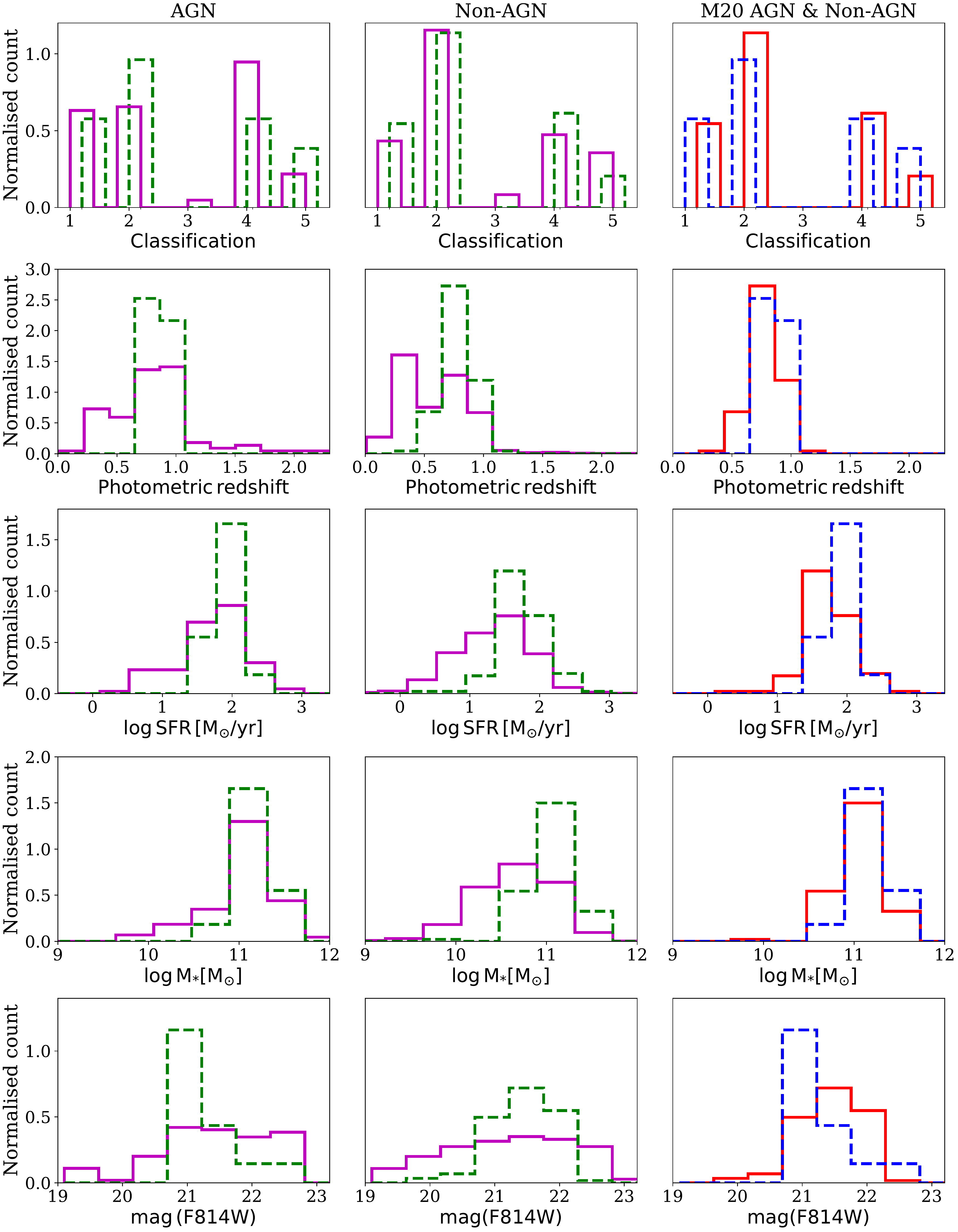}
\caption{\textit{Left column:} Normalised distributions of morphological types (first row), photometric redshift (second row), SFR (third row), stellar-mass (fourth row) and apparent F814W magnitude (bottom row) of AGN in the M20 subsample (dashed green histograms) and \protect\citetalias{Mahoro2019} sample (solid magenta histograms). \textit{Middle column:} Same as previous, but for non-AGN. \textit{Right column:} Same as previous, but comparing FIR AGN (blue-dashed lines) and non-AGN (solid red lines) of M20 subsample studied in this paper.}
\label{fig_parameters_comp}
\end{figure*}

In this section we compared normalised distributions of AGN and non-AGN green valley galaxies of the M20 sample analysed in this paper (see \autoref{fig_parameters_comp}, the right-hand column). First, we made a comparison based on morphological classification done in \cite{Mahoro2019}. We found a slight difference in class 2, where 38\% of AGN have spiral morphologies compared to 45\% of non-AGN (see \autoref{tab_Visual_number}). Secondly, we made a comparison of the photometric redshift, and found that AGN are at a slightly higher redshift, with 50\% of the sample being in the range of 0.69 and 0.93, while 50\%  of non-AGN green valley galaxies have redshifts in the range of 0.7 and 0.86. Thirdly, we compared the SFRs of AGN and non-AGN and found that AGN show higher SFRs, with 50\% of the sample being in a range of 63 $[M_{\odot}/yr]$ to 131.82 $[M_{\odot}/yr]$, while 50\% of non-AGN have SFRs between 33.88 $[M_{\odot}/yr]$ and 81.28 $[M_{\odot}/yr]$. Fourthly, we compared stellar mass (in $\rm{\log\, scale}$) of AGN and non-AGN without finding any significant difference, as can be seen in \autoref{tab_Visual_number}. Finally, we compared the F814W magnitudes and found that AGN are slightly brighter with 50\% of galaxies being in the range 21.17\,-\,21.45, in comparison to non-AGN being in the range 21.05\,-\,21.82.

In summary, our FIR AGN in the M20 sample are in average at a slightly higher redshift, with slightly higher SFRs, and a bit brighter in comparison to FIR non-AGN.

\section{STARLIGHT spectral fittings}\label{measurement_age}
In order to obtain the stellar population of AGN and non-AGN green valley galaxies, we model the stellar contribution in the LEGA-C spectra through the stellar population synthesis code STARLIGHT\footnote{STARLIGHT: http://www.starlight.ufsc.br/} \citep{CidFernandes2005}.
We checked our AGN spectra and found that all sources are type-2 AGN. Therefore, since we are not having type-1 AGN, and we are in addition masking all emission lines in spectra, we do not expect any significant AGN contribution to the continuum, as suggested in previous studies \citep[e.g.,][]{CidFernandes1998, Schmitt1999, Oio2019}. 

STARLIGHT fits an observed spectrum $O_\lambda$ with a model $M_\lambda$ that blends up to $N_{\star}$ single stellar populations (SSPs) with distinct ages and metallicities from different stellar population synthesis models. The code works out on the following equation for a model spectrum $\rm{M_{\lambda}}$:
\begin{equation} M_{\lambda} = M_{\lambda_{0}} \left[ \displaystyle
\sum_{j=1}^{N_{\star}} x_{j} b_{j,\lambda} r_{\lambda} \right]
\otimes G(v_{\star},\sigma_{\star}),
\end{equation}
where $b_{j,\lambda}\,r_{\lambda}$ is the reddened spectrum of the $j$-th SSP normalised at $\lambda_0$; $r_{\lambda}=10^{-0.4(A_{\lambda}-A_{\lambda 0})}$ is the reddening term; $M_{\lambda 0}$ is the synthetic flux at the normalisation wavelength; and $\vec{x}$ is the population vector. The sign $\otimes$ indicates the convolution operator and $G(v_{\star},\sigma_{\star})$ is the Gaussian distribution used to model the line-of-sight stellar motions. It is centered at velocity $v_{\star}$ with dispersion $\sigma_{\star}$. The addopted normalisation wavelength range of $\rm{4010\,\AA\,\leq\,\lambda_0\,\leq\,4160\,\AA}$ for observed spectra and at  the base of $\rm{\lambda_{0}\,=\,4020\,\AA}$. The best fit model is determined by minimising $\rm{\chi^2}$ (through a simulated annealing plus Metropolis scheme and Markov Chain Monte Carlo techniques):
\begin{equation}
\chi^2 = \sum_{\lambda}[(O_{\lambda}-M_{\lambda})w_{\lambda}]^2,
\end{equation}
where emission lines, bad pixels and unknown features are masked by assigning $w_{\lambda}$\,=\,0 to those regions.  We first moved to the rest-frame all observed spectra and corrected them for foreground Galactic extinction using the pystarlight\footnote{https://pypi.python.org/pypi/PySTARLIGHT} library and \cite{Schlegel1998} maps of dust IR emission. For the intrinsic extinction we used the \cite{Cardelli1989} law \citep[e.g.,][]{CidFernandes2011, Povic2016, Werle2019, Wild2020}.  We fitted the wavelength range from $\rm{3092\,\AA}$ to $\rm{6534\, \AA}$, that in most of cases present minimum and maximum wavelengths in our spectra.

The spectral used basis consists of 25 elements set, which is a sample of the Evolutionary Population Synthesis models given by \citet{Bruzual2003}. It covers 25 ages of 0.001, 0.00316, 0.00501, 0.00661, 0.00871,
0.01, 0.01445, 0.02512, 0.04, 0.055, 0.10152, 0.16090, 0.28612, 0.50880, 0.90479, 1.27805, 1.434, 2.5, 4.25, 6.25, 7.5, 10.0, 13.0, 15.0 and 18.0\,Gyr to provide enough resolution in age and solar metallicity which were computed with the \cite{Salpeter1955} initial mass function. The reason why we chose \cite{Bruzual2003} is that it has a wide range of wavelengths and it is also widely used in the literature, which allows us to compare our results and previous works directly. Furthermore, the \cite{Bruzual2003} is the base model for STARLIGHT, so it is convenient to use them together.\\
Nevertheless, to test the robustness of results, we fitted a random selection of 10 AGN and 10 non-AGN using both \cite{Bruzual2003} model and Granada-Miles (GM) Single Stellar Population explained in \cite{CidFernandes2013, CidFernandes2014} that is built using the MILES \citep{Vazdekis2010} and \cite{GonzlezDelgado2005} models.
In general, we obtained similar results when using the \cite{Bruzual2003} and GM models, with differences $\le$\,10\% in intermediate and old stellar populations. Therefore, we do not expect any significant uncertanities in our results due to the use of \cite{Bruzual2003} model. In addition, we do not expect any AGN contribution to the continuum, since having type-2 AGN sources as mentioned above, but nevertheless, we also measured stellar populations and ages after adding a non-stellar component during the
fitting process, without finding any significant differences in our results.

In this work, we assumed solar metallicity (Z\,=\,0.02), in line with previous studies for galaxies of similar stellar masses \citep[e.g.,][]{ValeAsari2009, Povic2016}. However, we also tested the use of a mixture of three different metallicities: sub-solar (Z\,=\,0.008), solar (Z\,=\,0.02), and super-solar (Z\,=\,0.05). In general, we do not find significant differences when using solar and a mixture of metallicities, with solar metallicity giving a slightly higher light-weighted mean ages by 0.14 dex in AGN and 0.07 dex in non-AGN. 

We made use of the \textit{adev} parameter which gives goodness of the fit and presents the average deviation over all fitted pixels (in percentage). Fitting with \textit{adev} $<$ 6 yields very good fits, and such cases were selected for analysis \citep {CidFernandes2005, Povic2016}. For those spectra which had given \textit{adev} in the range 6 $\leq$ \textit{adev} $<$ 10, we applied a three-pixel boxcar smoothing to the spectra before re-running STARLIGHT. We then selected also those spectra with \textit{adev} $<$ 6  after this second STARLIGHT fitting for analysis, after also visually checking that all such cases had resulted in good and stable fits. However, for consistency, for the spectral stellar population analysis, we used the SP output obtained by the first STARLIGHT run (i.e. the fit without smoothing, though we note that there were no significant differences in the SP results with or without smoothing).

\begin{figure*}
\centering
\begin{minipage}[c]{.49\textwidth}
\includegraphics[width=0.89\textwidth,angle=0]{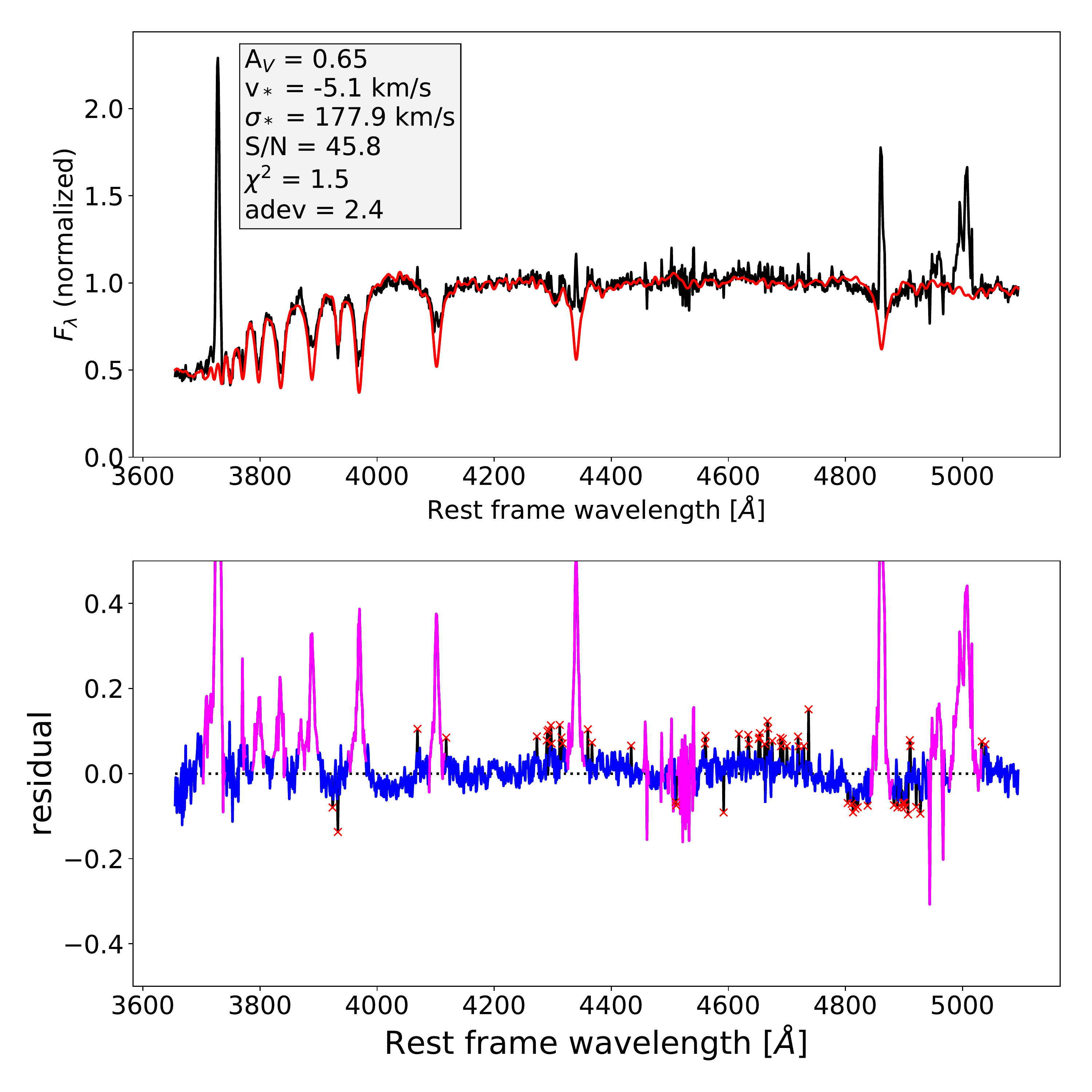}
\end{minipage}
\begin{minipage}[c]{.49\textwidth}
\includegraphics[width=0.89\textwidth,angle=0]{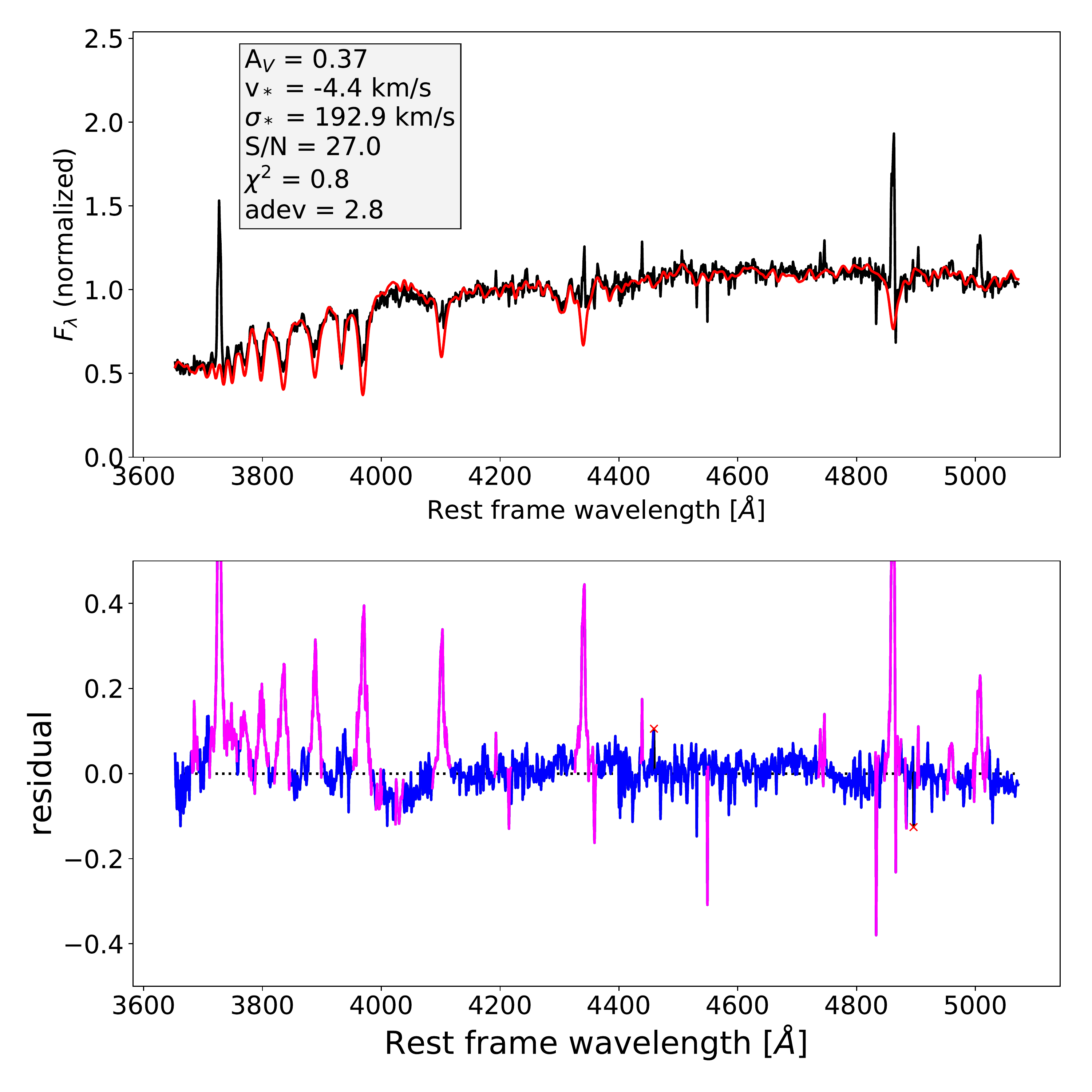}
\end{minipage}
\caption[ ]{Example of STARLIGHT spectral fitting of AGN (left plot) and non-AGN galaxy (right plot). The top panel shows the data ($\rm{O\lambda}$, black line) and STARLIGHT fit ($\rm{M\lambda}$, red line). The bottom panel presents the $\rm{O\lambda\,-\, M\lambda}$ residual spectrum. Magenta lines correspond to masked windows.} 
\label{starlightout}
\end{figure*}
In \autoref{starlightout}, we show some examples of the best STARLIGHT fits (red lines) and original spectra (black lines). The code output gives the `population vector' $(x_{j})$, which is the fraction of the total light that each simple stellar populations (SSP) contributes to the fitting.
The total sum of $x_{j}$ vectors from the SFH solution in STARLIGHT might be larger than 100\%. We need to normalise these vectors if we want to use them to obtain weights for estimating mean properties. In this way, $x_{j}$ vectors were normalised to 100\% before we measured the light-weighted mean age.
We derived the light-weighted mean age from the spectrum as a function of $x_{t}$ through:
\begin{center}
\label{eq_age_starlight}
$\rm{\left <\,\log\,t\right >\,=\,\sum\limits_{t,Z}\,x_{t,Z}\log\,t}$,
\end{center}
where $x_{t, Z}$ is a fraction of light at stellar age t in our best-fitting model and metallicity Z (which is fixed in our case as Z$_{\odot}$). The derived ages are a 'light-weighted' mean of the total light.

To obtain a general view of the stellar populations of AGN and non-AGN, we arranged their stellar populations into three groups: old with age $\rm{>\,10^{9}\,yr}$, intermediate with age between $\rm{10^{8}\,yr\,<\,age\le\,10^{9}\,yr}$, and young populations with age $\rm{\le\,10^{8}\,yr}$, in line with previous studies \citep[e.g.,][]{CidFernandes2005,  CidFernandes2013, Povic2016, Cai2020}.

\section{Analysis and results}
\subsection{Stellar populations of FIR AGN and FIR non-AGN green valley galaxies}\label{stellar_population}
\begin{figure}
\centering
\includegraphics[width=\columnwidth]{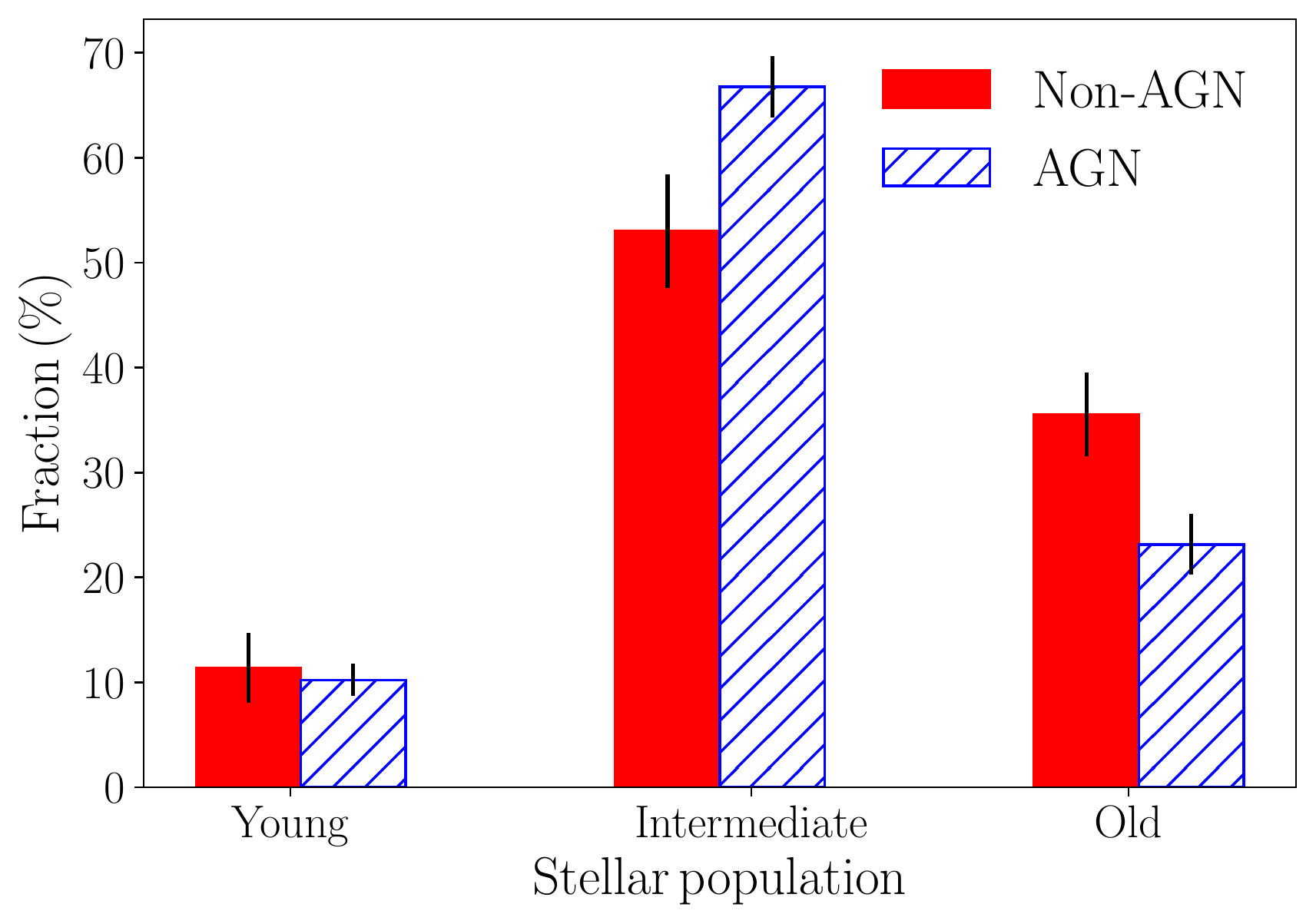}
\caption{Stack bar plot showing the fractional number of FIR AGN (blue hatched) and non-AGN (red solid) green valley galaxies whose average age falls into the three defined ranges: young (age [yr] $\le$ 10$^{8}$), intermediate (10$^{8}$ $<$ age [yr] $\le$ 10$^{9}$), and old (age [yr] $>$ 10$^{9}$).}
\label{fig.age.stellar.population}
\end{figure}

In this section, we present the obtained stellar populations using the data described in \Cref{specdata} and methodology given in \Cref{measurement_age}. \autoref{fig.age.stellar.population} shows the fraction of young, intermediate, and old stellar populations of our FIR AGN (in blue hatched) and non-AGN (in red solid) green valley galaxies. In general, we find that both AGN and non-AGN green valley galaxies are dominated by intermediate stellar populations. We can also see that AGN have a slightly higher fraction of intermediate stellar populations in comparison to non-AGN (67\% vs. 53\%, respectively) and a lower fraction of old stellar populations (23\% vs. 36\%, respectively). This result goes in line with results previously obtained in \cite{Mahoro2017, Mahoro2019}, where we found that FIR green valley galaxies are predominantly located on the MS of SF regardless of whether they are AGN or non-AGN, and that FIR green valley AGN show higher SFRs. In addition, we compared a sub-sample of 13 AGN and 69 non-AGN green valley galaxies to match the same total stellar mass range of 10.8\,$\le$\,logM*\,$\le$\,11.6 and redshift range of 0.68\,$\le$\,z\,$\le$\,1.05. We obtained very similar stellar populations of the two sub-samples when compared to the total M20 AGN and non-AGN samples, with differences  $<$\,10\%. Therefore, with our spectral synthesis analysis we do not observe that our AGN have older stellar populations in comparison with non-AGN, as has been suggested in previous optical studies \citep[e.g.,][]{Leslie2016}. Finally, only a small fraction ($\rm{\sim}$10\%) of both AGN and non-AGN show young stellar populations, which is again in line with previous studies of green valley galaxies \citep[e.g.,][and references therein]{Angthopo2020}.   

To test the robustness of our results, we used the same methodology as used by \cite{Dametto2014}. To estimate the uncertainties in our data, we selected randomly 7 AGN and 7 non-AGN green valley galaxies and then perturbed them with the error spectra in a Gaussian random distribution to measure the error. This gave us a new spectrum disturbed by the error, that is, a new realisation of the spectrum. Then we created 100 spectra, run again STARLIGHT, and then recovered each property. The main properties derived from the spectral synthesis are light weighted mean ages and stellar populations. However, before analysing the spectra, it is helpful to assess uncertainties in the individual spectra as if they came from different galaxies. We found average age uncertainties in analysed AGN (non-AGN) of 2\% (3\%), 3\% (5\%) and 3\% (4\%) in the case of the young, intermediate, and old ages, respectively. We found that obtained average uncertainties are in line with previous studies \citep[e.g.,][]{CidFernandes2014, Burtscher2021, DahmerHahn2022}. The measured averaged uncertainties are included in \autoref{fig.age.stellar.population}.

\subsection{Stellar ages}\label{Stellar_ages}
\begin{figure}
\centering
\includegraphics[width=\columnwidth]{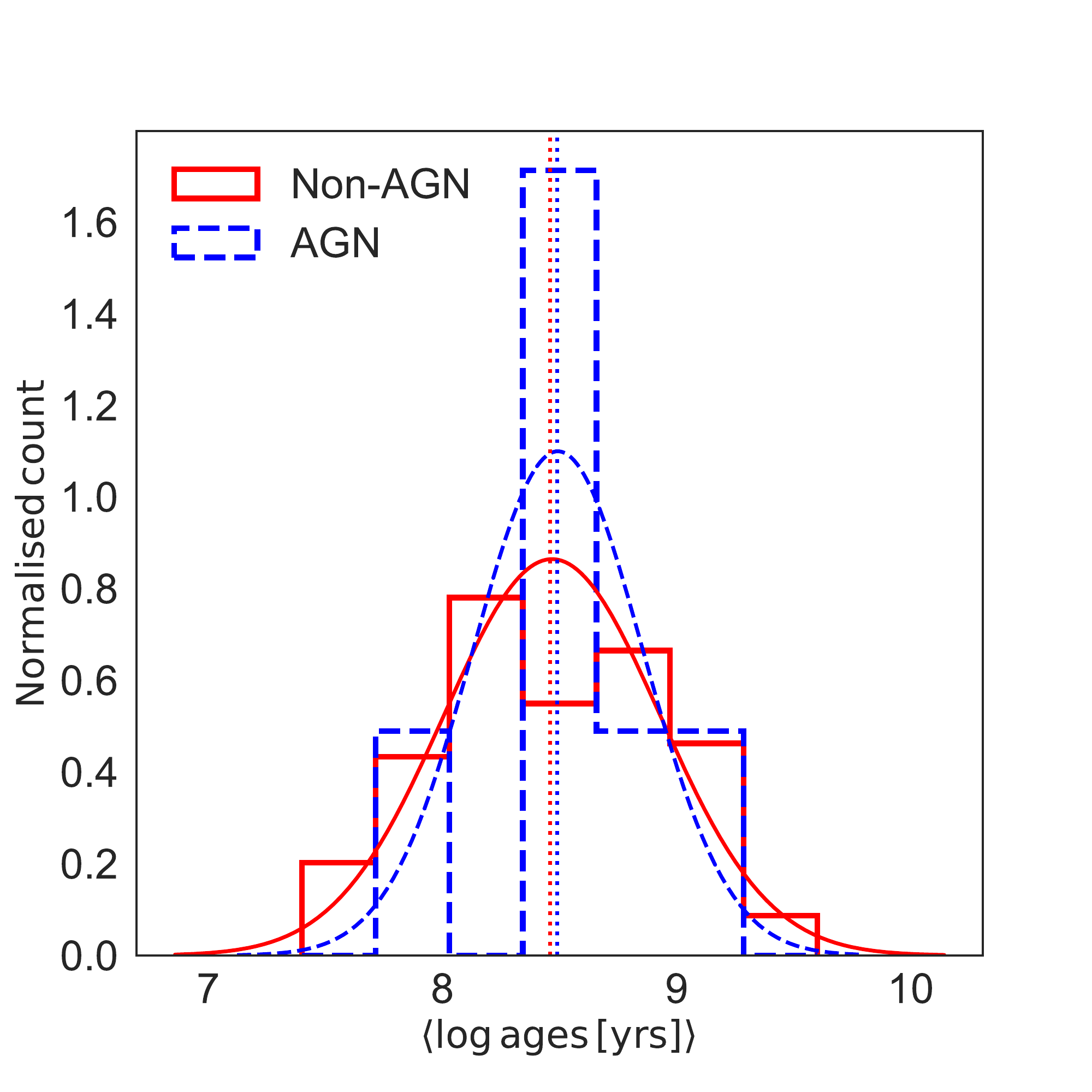}
\caption{Normalised distribution of light-weighted mean ages of FIR green valley AGN (blue dashed line) and non-AGN (red solid line) galaxies with  the respective Gaussian that best fits the distributions. The vertical dashed lines represent the median values for each histogram, where each median has the same colour as one of the corresponding histograms.}
\label{ages}
\end{figure}

In this section we compare the light-weighted mean ages of FIR AGN and non-AGN green valley galaxies measured from STARLIGHT output, as explained in \Cref{measurement_age}. \autoref{ages} shows the normalised distributions of the light-weighted mean age in the M20 sample, along with best-fit Gaussian functions fitted to the two distributions, highlighting their similarity. In line with previous section, AGN and non-AGN span a similar range of ages. The median age of AGN is $\rm{\log t\,=\,8.5\,[yr]}$, with 50\% of sources covering the range of $\rm{\log t\,=\,8.4\,\--\,8.6\,[yr]}$, while the median age of  non-AGN is $\rm{\log t\,=\,8.4\,[yr]}$, with 50\%  having ages in the range of $\rm{\log t\,=\,8.2\,\--\,8.9\,[yr]}$. Similar stellar ages are also obtained when comparing the AGN and non-AGN subsamples within the same stellar mass and redshift ranges, as indicated in the previous section.

\begin{figure*}
\centering
\begin{minipage}[c]{.49\textwidth}
\includegraphics[width=0.89\textwidth,angle=0]{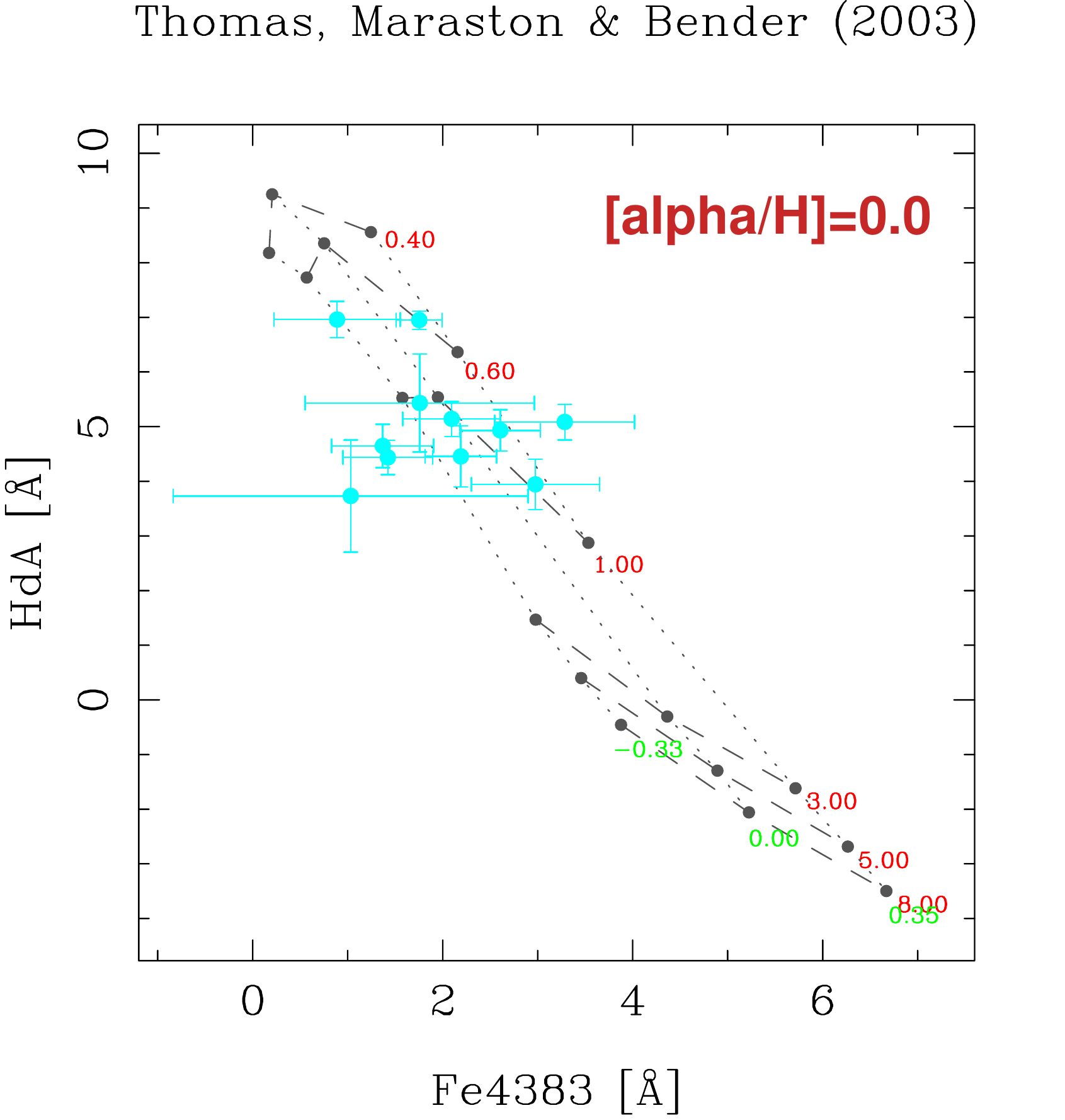}
\end{minipage}
\begin{minipage}[c]{.49\textwidth}
\includegraphics[width=0.89\textwidth,angle=0]{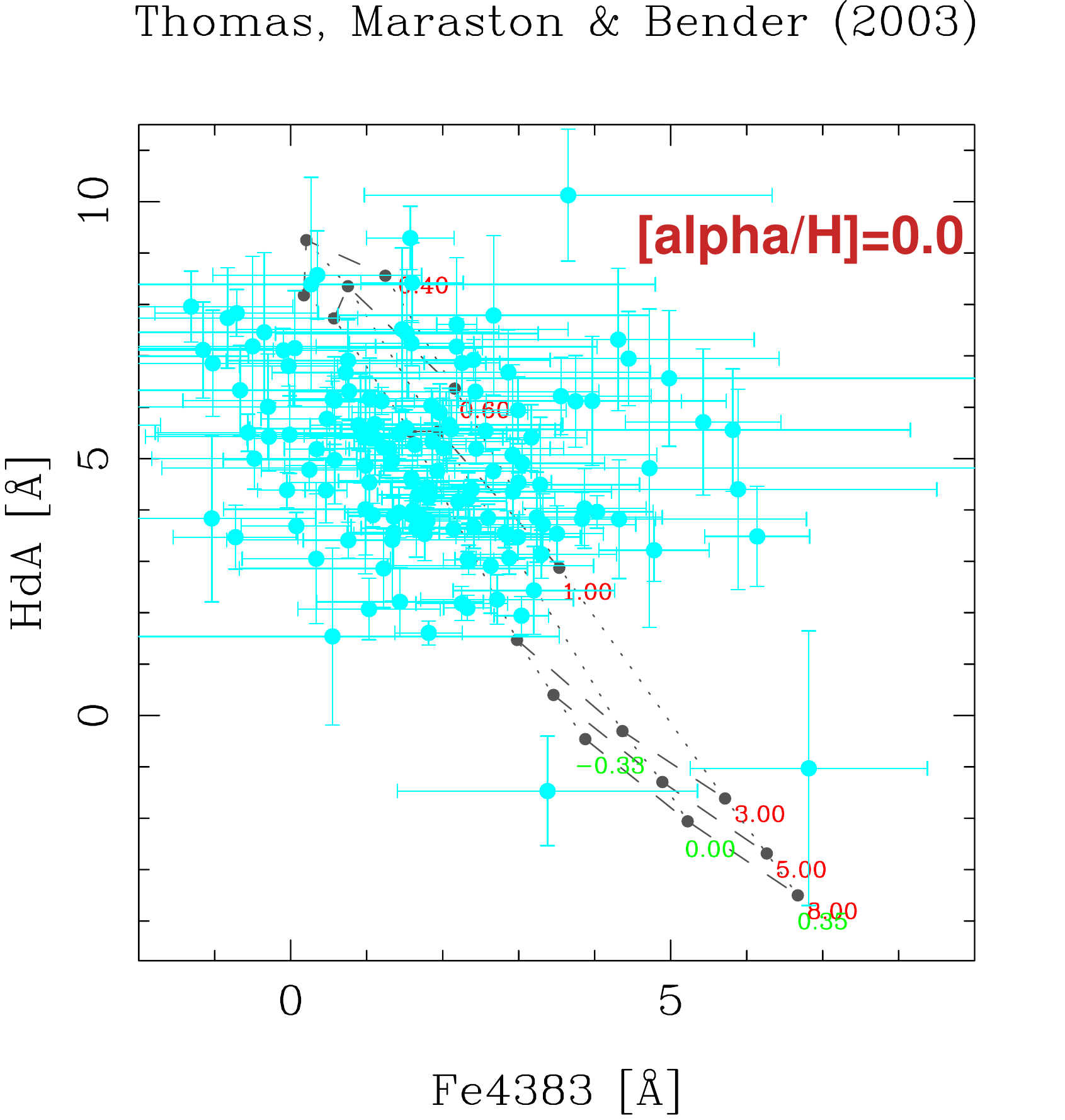}
\end{minipage}
\caption{H$\rm{\delta_{A}}$ vs. Fe4383 diagram for FIR AGN (left) and FIR non-AGN (right) green valley galaxies plotted over the \protect\cite{Thomas2003} models. The model grid is the same in both plots. Dashed and dotted lines of the grid correspond to the predictions for constant age (denoted by red numbers in Gyr) and metallicity (green numbers), respectively. } 
\label{FE4383_HD_A_alpha_00}
\end{figure*}

Since in STARLIGHT fittings we assumed a fixed solar metallicity (see \Cref{measurement_age}), we did an independent and additional check of our analysis using the “\textit{rmodel}”\footnote{https://rmodel.readthedocs.io/en/latest/index.html} code \citep{Cardiel2003}. This code uses different SSPs \citep{Bruzual2003, Thomas2003, Vazdekis2003, Lee2005} and determines the stellar populations' parameters, such as age, metallicity, the slope of the initial mass function, etc., using a pair of line intensity indices as an input. We ran the \textit{rmodel} using the \cite{Thomas2003} SSPs. \autoref{FE4383_HD_A_alpha_00} shows the positions occupied by our sources in H$\rm{\delta_{A}}$ vs. Fe4383 parameter space (the $\rm{[\alpha/Fe]}$ parameter was fixed to 0.0) taken from the public catalogue of \cite{Straatman2019}. It can be seen that the locus of distribution in \autoref{FE4383_HD_A_alpha_00} as far as metallicities are concerned (green numbers of the grid) aligns with results obtained by STARLIGHT in \Cref{ages}, where we assumed solar metallicity. 

However, the locus of the distribution of points in \autoref{FE4383_HD_A_alpha_00} indicates stellar ages of approximately 1\,Gyr (the red figures of the grid).  This apparent discrepancy is explained by the fact that the absorption line indices probe mass-weighted ages, rather than light-weighted ages.  We thus also measured the mass-weighted stellar ages using the STARLIGHT output. The obtained mass-weighted stellar ages cover a range of $\rm{\log t\,=\,8.8\,-\,9.6\,[yr]}$ and $\rm{\log t\,=\,8.8\,-\,9.8\,[yr]}$ for FIR AGN and non-AGN, respectively, with a median age of $\rm{\log t\,=\,9.1\,[yr]}$ in both cases. Therefore, it is clear that mass-weighted stellar ages of our FIR green valley galaxies is in good alignment with the absorption line indices analysis performed with \textit{rmodel} code. However, throughout the rest of the paper our analysis focuses on light-weighted stellar ages.

\subsection{Dn(4000) vs. H$\rm{\delta_{A}}$ diagram}\label{Dn4000vsHA}

Furthermore, we used the 4000 $\angstrom$ break (Dn4000) and the Balmer absorption-line index H$\rm{\delta_{A}}$ available in the catalogue of \cite{Straatman2019} to study the M20 sample of green valley galaxies. These two indices are essential for tracing the star formation histories (SFHs) in galaxies \citep[e.g.,][]{Kauffmann2003}. The spectral indices Dn4000 and H$\rm{\delta_{A}}$ \citep{Balogh1999} were derived following the analysis in \cite{Gallazzi2014}, and the measurements are fully explained in \cite{Straatman2018}.

\begin{figure*}
\centering
\begin{minipage}[c]{.49\textwidth}
\includegraphics[width=0.89\textwidth,angle=0]{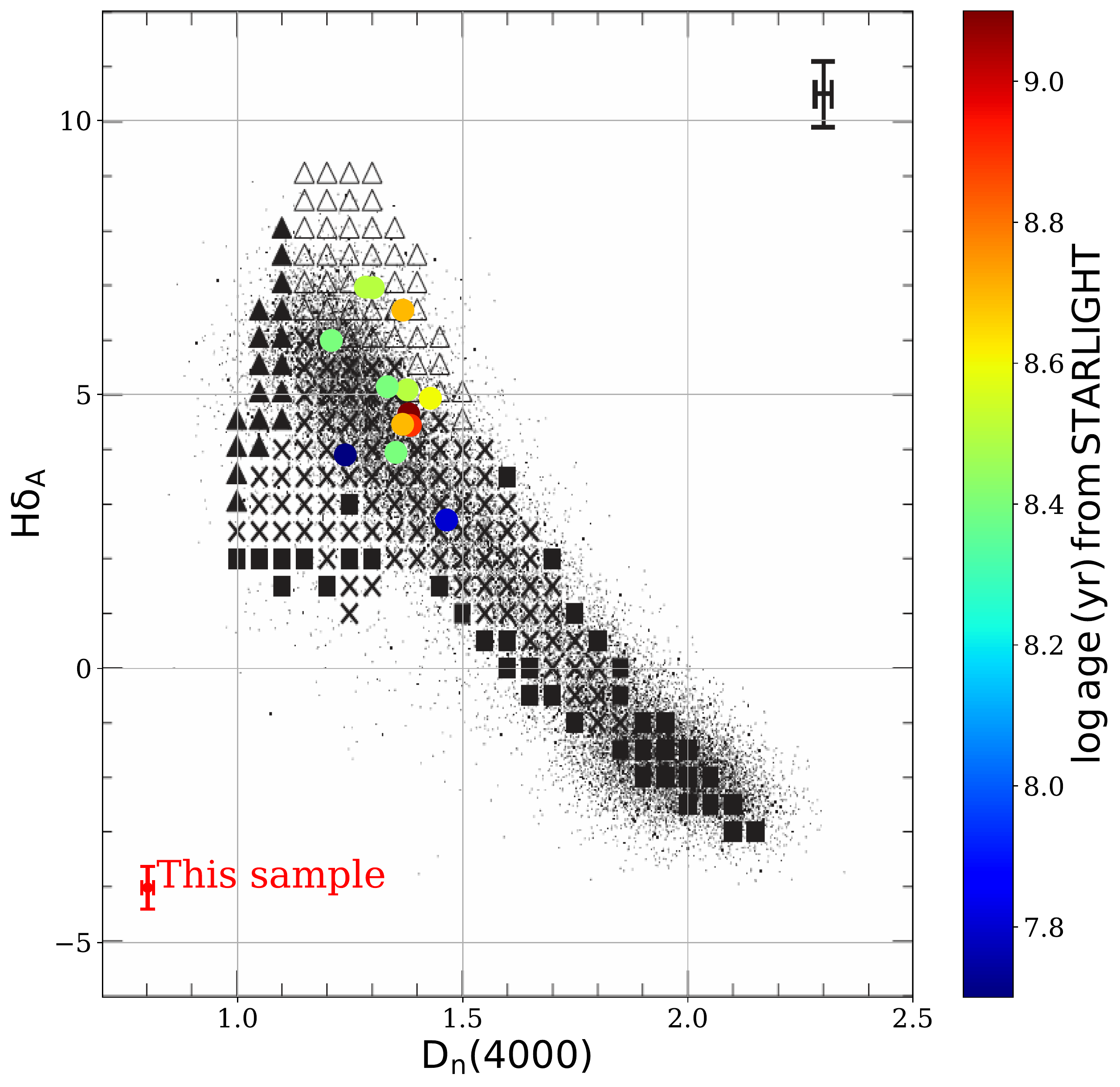}
\end{minipage}
\begin{minipage}[c]{.49\textwidth}
\includegraphics[width=0.89\textwidth,angle=0]{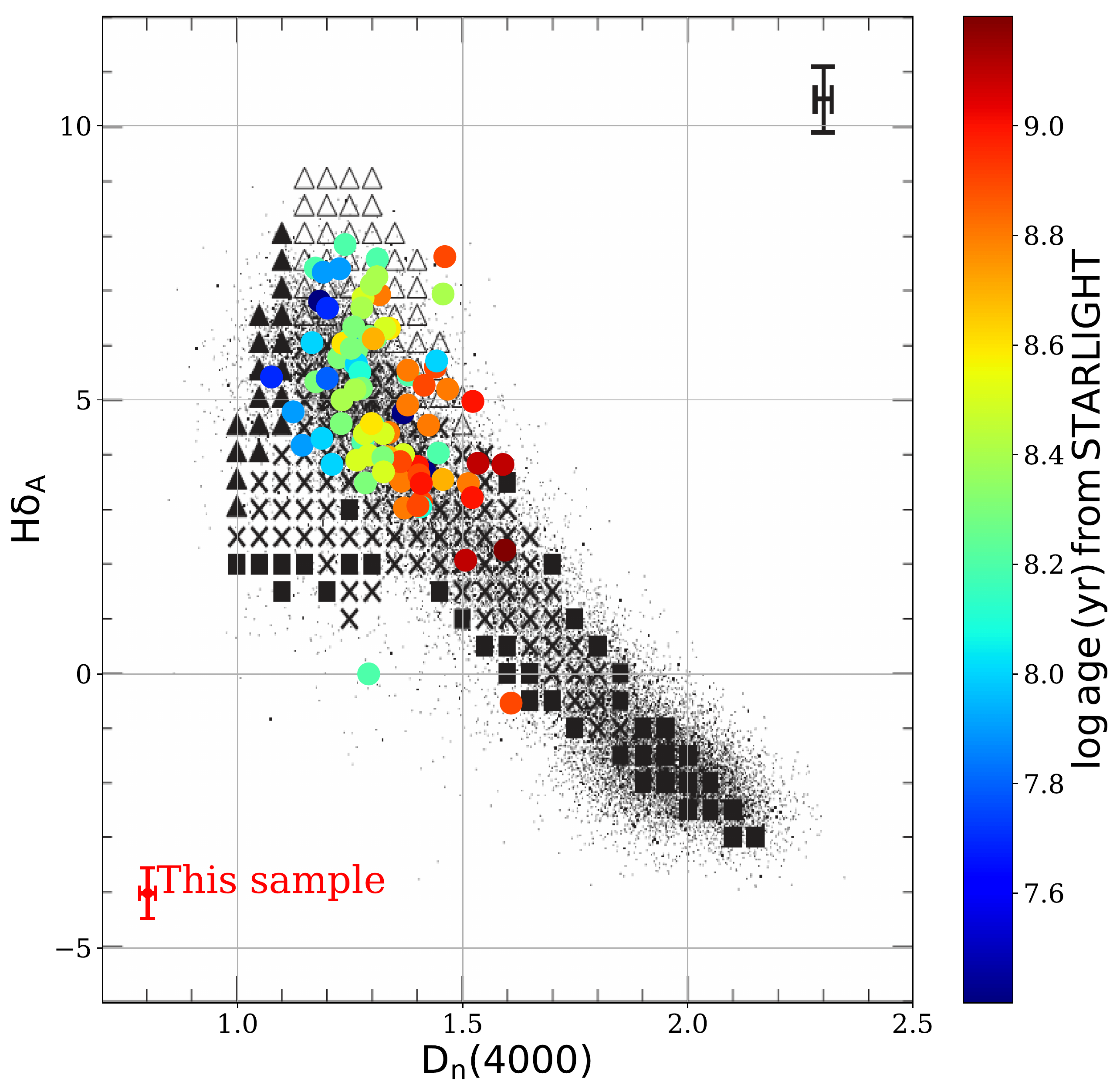}
\end{minipage}
\caption{Figure taken from ~\protect\cite{Kauffmann2003} (see their Figure 3) showing the relation between Dn4000 and H$\rm{\delta_{A}}$ indices for their sample of SDSS galaxies (dots) with modelled indices (larger symbols). Regions with triangles show areas where a galaxy has experienced a burst over the past 2\,Gyr with high confidence, open triangles show older bursts of more than 0.1\,Gyr of age, and closed triangles show more recent bursts of less than 0.1\,Gyr of age. Solid squares indicate regions where 95\% of model galaxies have F$_{\rm{burst}}$ = 0, i.e. have continuous SF.  Regions marked with crosses contain a mix of burst and continuous star formation models. Finally, overlaid by coloured circular markers is our sample of AGN (left) and non-AGN (right), where the colour code corresponds to stellar ages derived by STARLIGHT, as indicated in the legend.} 
\label{HDA_Dn4000}
\end{figure*}

\autoref{HDA_Dn4000} shows the relation between the Dn4000 and H$\rm{\delta_{A}}$ indices obtained by \cite{Kauffmann2003} for the SDSS DR4 sample (see their Figure 3). The authors used a library of 32000 different star formation histories of stellar populations, where for each star formation history, there are corresponding Dn4000 and H$\rm{\delta_{A}}$ indices, as well as the fraction of total stellar mass of the galaxy formed in the bursts over 2 Gyr (F$_{\rm{burst}}$). We used \cite{Kauffmann2003} Figure 6, where bins are coded according to the fraction of modelled SFHs with F$_{\rm{burst}}$ in a given range, and overlaid our AGN and non-AGN green valley galaxies (coloured filled circles in \autoref{HDA_Dn4000}) using the Dn4000 and H$\rm{\delta_{A}}$ measurements of \cite{Straatman2018}. The colouring scheme is used to represent the age derived from our STARLIGHT output. All analysis of the relation between Dn4000 and H$\rm{\delta_{A}}$ was conducted using the AGN and non-AGN samples selected in the same stellar mass range of $\rm{10.6\,M_{\sun}\,-\,11.6\,M_{\sun}}$, as explained in \cite{Mahoro2017}. When comparing Dn4000 and H$\rm{\delta_{A}}$ between AGN and non-AGN samples, very small or no difference has been found. In the case of Dn4000 index, AGN and non-AGN have a median value of 1.36 and 1.30, respectively, while the median values of H$\rm{\delta_{A}}$ index are 4.9 and 4.7, respectively.\\
\indent Given that the colour of the circular markers in Dn4000 vs. H$\rm{\delta_{A}}$ plot in \autoref{HDA_Dn4000} is an indicator of age, we can visually evaluate the consistency of the age determination from the two methods. From \autoref{HDA_Dn4000}, we can see in the case of non-AGN a gradual change from blue (young) to red (old) colours as Dn4000 increases, as suggested in previous studies \citep[e.g.,][]{Kauffmann2003, HernnCaballero2013, Siudek2017, Wu2018}. The trend, however, is not clear for AGN, but it may be affected by the small number of sources.\\  
\indent From the location of the M20 sample in \autoref{HDA_Dn4000}, it can be seen that we have significantly more sources showing a mixture of both burst and continuous star formation, rather than pure bursts, which is in line with results obtained in \cite{Mahoro2017, Mahoro2019} where most of AGN and non-AGN are located on and above the MS of SF.
Following the \cite{Kauffmann2003} models, we obtained a similar trend in both samples where most of the sources (62\% of AGN and 66\% of non-AGN) could have experienced both bursts and continuous SF. These results are in line with measurements of stellar populations and ages obtained in \Cref{stellar_population} and \Cref{Stellar_ages}. 

\indent Finally, an insignificant fraction of our AGN and non-AGN are falling in the young $<$\,0.1\,Gyr burst region. However, a slightly higher fraction of AGN (38\%) compared to non-AGN (27\%) might have experienced a burst of SF earlier, more than 0.1\,Gyr ago, again in line with results obtained in \cite{Mahoro2017, Mahoro2019} where we found larger SFRs for FIR AGN.

\subsection{Main sequence of star formation}\label{mainsequence}
One of the ways to study galaxy evolution is to analyse the relation between SFR and stellar mass to check their location in comparison to the MS of star-forming galaxies \citep[e.g.,][]{Elbaz2007, Whitaker2012, Davies2016, Leslie2016, Povic2016, Elbaz2018, Popesso2019, Nantais2020}. The details on SFR and stellar mass measurements are explained briefly in \Cref{data}, with more details given in \cite{Mahoro2017}.

\begin{figure*}
\centering
\includegraphics[width=16cm]{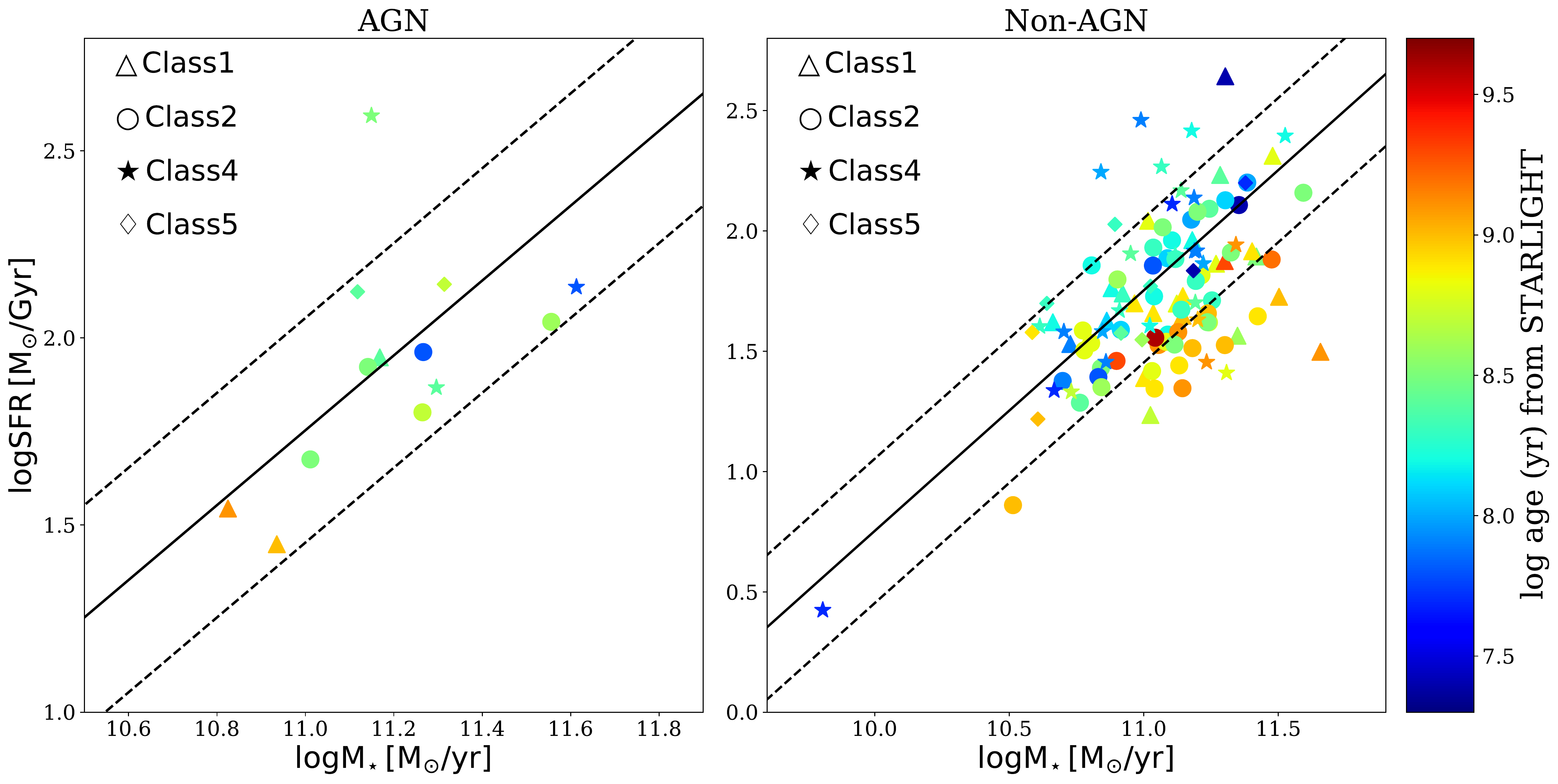}
\caption{SFR vs. stellar mass diagram for our sample of FIR green valley AGN (left) and non-AGN (right) green valley galaxies. The MS of star-forming galaxies \citep{Elbaz2011} and $\rm{1\,\sigma}$ range are given with solid and dashed lines, respectively. The symbols are colour coded by the light-weighted stellar ages from STARLIGHT, while the symbol type depicts the galaxies' visual morphological class from ~\protect\cite{Mahoro2019}, where class 1 are elliptical, S0 or S0/S0a galaxies, class 2 are spirals, class 4 are peculiar galaxies, and class 5 are unclassified galaxies. We do not have in our M20 sample galaxies classified as class 3 (irregular) in ~\protect\cite{Mahoro2019}.}
\label{MSage}
\end{figure*}

The positions of M20 AGN and non-AGN green valley galaxies in the $\rm{SFR-M_{*}}$ plane, as derived in \cite{Mahoro2017, Mahoro2019}, are shown in \autoref{MSage}, taking now into account measured stellar ages. All our galaxies are coloured according to their light-weighted mean age, while the type of the symbol indicates their morphological classification from \cite{Mahoro2019}.
For the MS of SF, we used the fit obtained by \cite{Elbaz2011} whose SFRs are based on FIR \textit{Herschel} data. For the width of the MS we used $\rm{\pm\,0.3}$ dex (dashed lines), found in many previous works to be the typical $\rm{1\,\sigma}$ \citep[e.g.,][]{Elbaz2007,Whitaker2012, Whitaker2014, Shimizu2015, Povic2016, Santini2017, Donnari2019, CurtisLake2021}.
\autoref{MS_fraction} provides the number of galaxies above, on, and below the MS of SF. For the non-AGN (right plot) we can see that galaxies with younger stellar ages are located mainly on the MS of SF, while those with older ages are found below the MS, as expected. This trend, however, can not be clearly seen for AGN sources, where almost all sources are located on the MS independently on their stellar ages. Finally, \autoref{Morph_stellar} summarises the number of galaxies in class1, class2, class4 and class5 in relation to stellar ages. We do not have in our M20 sample galaxies classified in \cite{Mahoro2019} as class 3 (irregular). In general, we do not find significant differences between AGN and non-AGN regarding the fractions of stellar ages of different morphological types.

\begin{table}
\centering
\caption{Number and fraction of AGN and non-AGN with respect to the MS of SF.}
\label{MS_fraction}
\begin{tabular}{cccc} \hline
        & Above MS & On MS     & Below MS  \\
AGN     & 1 (8 \%)  & 12 (92 \%) & -         \\ 
\hline
Non-AGN & 7 (6\%)   & 81 (74 \%) & 22 (20\%)  \\
\hline
\end{tabular}
\end{table}

\begin{table}
\caption{Summary of morphological types in relation with stellar ages.}
\label{Morph_stellar}
\begin{tabular}{cccc|ccc}
\multicolumn{1}{l}{} & \multicolumn{3}{c|}{AGN} & \multicolumn{3}{c}{Non-AGN} \\ \cline{1-7} 
\multicolumn{1}{l}{} & Young   & Inter    & Old        & Young    & Inter    & Old   \\ \cline{1-7} 
Class1               & 0 (0\%) & 2 (15\%) & 1 (8\%)    & 2 (2\%)        & 20 (18\%)      & 2 (2\%)  \\
Class2               & 1 (8\%) & 4 (31\%) & 0 (0\%)    & 6 (5\%)       & 37 (34\%)      & 7 (6\%)    \\
Class4               & 1 (8\%) & 2 (15\%) & 0 (0\%)    & 12(11\%)       & 13 (12\%)      & 2  (2\%)   \\
Class5               & 0 (0\%) & 2 (15\%) & 0 (0\%)    & 2 (2\%)       & 7 (6\%)        & 0  (0\%)  
\end{tabular}
\end{table}

\subsection{Star formation time-scales}\label{time_scales}
A constant star formation rate describes the star formation histories in galaxies for the first few Gyr since their birth (possibly characterised by burst), typically followed by a period of exponentially declining star formation rate \citep[e.g.,][]{Martin2007, Goncalves2012, Schawinski2014, Hahn2017, NogueiraCavalcante2018, NogueiraCavalcante2019, Carleton2020}. In this section, we measure the SF time-scale ($\rm{\tau}$) using SFR(t) $\sim$ exp(-t/$\rm{\tau}$), where t is time (i.e., our light-weighted measured stellar age), SFR is as measured in \cite{Mahoro2017, Mahoro2019}, and the exponential index ($\rm{\tau}$) is a time-scale in units of $\rm{Gyr^{-1}}$ \citep[e.g.,][]{Bell2000, Taylor2011}. Using the above equation, we obtained similar median time-scales for FIR AGN and non-AGN green valley galaxies of $\rm{\sim}$ 71\,Myr and $\rm{\sim}$ 69\,Myr, respectively. In addition, 50\% of our AGN and non-AGN have time-scale values in a range of 53 Myr - 103 Myr and 36 Myr - 194 Myr, respectively. More discussion on quenching time-scales is given in \Cref{time_scale_discuss}.

\section{Discussion}
In general, galaxies are composed of multiple stellar populations of different ages, and stellar populations are some of the key parameters for understanding the properties of galaxies and their evolution. This paper has aimed to understand the stellar population properties of AGN and non-AGN galaxies with FIR emission in the green valley, and to compare the obtained results with those obtained previously in \cite{Mahoro2017, Mahoro2019} where enhanced SFRs have been found in AGN host galaxies, in contradiction with previous results carried out in optical. 
\Cref{stellar_population} and \Cref{Stellar_ages} presented comparisons of stellar populations and stellar ages of AGN and non-AGN green valley galaxies at redshifts of 0.5 $\rm{<}$ z $\rm{<}$ 1.0), showing that their stellar populations are dominated by intermediate stellar ages. Our results are consistent with \cite{Pan2013}, who examined green valley galaxies using COSMOS field data at 0.2 $\rm{<}$ z $\rm{<}$ 1.0, and found intermediate stellar populations and intermediate stellar ages. Our results are also in agreement with a lower redshift sample of \cite{Phillipps2019}, who used the Galaxy and Mass Assembly (GAMA) data at 0.1 $\rm{<}$ z $\rm{<}$ 0.2 finding that green valley galaxies show intermediate stellar population ages. Moreover, \cite{Angthopo2020} used Sloan Digital Sky Survey (SDSS) data with green valley being defined through the 4000 $\rm{\AA}$ break strength, finding again that green valley galaxies are characterised by intermediate stellar ages, in line with \cite{Trussler2020} results when using SDSS DR7 galaxies in the redshift range of 0.02 $\rm{<\,z\,<}$ 0.085. 

In this work, we confirm that intermediate stellar populations and intermediate stellar ages dominate also in the case of FIR selected green valley galaxies, both AGN and non-AGN. In addition, in \Cref{Dn4000vsHA} we investigated the distribution of our sample in the Dn4000 vs. H$\rm{\delta_{A}}$ diagram, and found that a high fraction of both AGN (62\%) and non-AGN (66\%) show evidence of both bursts and continuous SF in their star-formation histories. Our findings suggest that both AGN and non-AGN green valley galaxies in our sample are star-forming galaxies, which explains their location on the main sequence of SF. It is also important to note that our M20 sample shows FIR emission, and that our Dn4000 median value is in line with the one obtained in \cite{AlonsoHerrero2010} for local luminous infrared galaxies at $\rm{z\,<\,0.26}$.

The SFR vs. $\rm{M_{*}}$ diagram (SFR-$\rm{M_{*}}$) has been broadly explored in the literature \citep[e.g.,][]{Belfiore2018, Sanchez2018, Sanchez2019, Lacerda2020} and has served as an important diagnostic tool for studying galaxy evolution and transformation. As expected from their intermediate colours and physical properties, green valley galaxies in general lie mostly below the MS of star-forming galaxies \citep[e.g.,][]{Noeske2007, Salim2007, Renzini2015, Smethurst2015}. In our sample, however, most of the sources are located on the MS of SF, as can be seen in \autoref{MSage} and \autoref{MS_fraction} and in \cite{Mahoro2017, Mahoro2019}. For non-AGN we can observe a trend where the galaxies with older stellar ages have a tendency to be below the MS, as compared to those with intermediate and young stellar ages. These results are consistent with e.g., \cite{Ciesla2017}, who used simulations and star-forming galaxies in the GOODS-South field at $\rm{1.5\,<\,z\,<\,2.5}$, finding younger stellar ages on the MS and above, while older galaxies are located below MS. This trend, however, is not so clear in the case of our AGN sample, located mainly on and above the MS of SF, independently on their stellar ages. We need larger AGN spectroscopic samples to confirm this, however, results obtained in this paper go in line with those obtained in \cite{Mahoro2017, Mahoro2019}, where enhanced SFRs (measured in FIR) have been found in AGN host galaxies in comparison to non-AGN, within the same stellar mass range and independently on morphology. Therefore, stellar population analysis carried out in optical in this paper also suggests that FIR AGN in the green valley do not show clear signs of quenched star formation and negative AGN feedback, as suggested previously in different X-ray and optical studies \cite[e.g.,][]{Nandra2007, Povic2012, Leslie2016}.\,Moreover, recently it was also suggested that local AGN are not just a simple transition type between star-forming and quiescent galaxies, and that both star formation and AGN activity can be triggered simultaneously \citep{MartnNavarro2021}.

\subsection{FIR AGNs and non-AGNs quenching time-scales}\label{time_scale_discuss}

The intermediate nature of the green valley suggests that green valley galaxies are quenching their star formation, starting in the blue cloud and reaching the red sequence afterward. Some of the studies on the quenching time-scale of green valley galaxies have shown that the transition can occur in a short time-scale of less than 1 Gyr, while others have identified multiple transitional states in green valley galaxies and reported time-scales from intermediate (1–2 Gyr) to slow quenching \citep[e.g.,][]{Springel2005, Faber2007, Pan2013, Schawinski2014, Smethurst2015, Bremer2018, Belfiore2018, Phillipps2019, Angthopo2019}. 

Some of the studies suggested different quenching time-scales of early- and late-type galaxies. \cite{Schawinski2014} using a sample of SDSS+GALEX+Galaxy Zoo data at low redshift found that quenching time-scale of early-type galaxies ($<$\,250\,Myr) may be much shorter than that of late-type galaxies ($>$\,1\,Gyr). Similarly, \cite{NogueiraCavalcante2018} measured quenching time-scales in green valley galaxies at intermediate redshifts ($\rm{0.5\, {\ensuremath{\lesssim}}\,z\,{\ensuremath{\lesssim}}\,1.0}$) using the zCOSMOS data, finding lower star formation quenching time-scales in comparison to \cite{Schawinski2014}. They found that time-scales of green valley disks are $\rm{\geq}$ 250 Myr, and shorter values of 150\,Myr, 100\,Myr and 50\,Myr for elliptical galaxies, irregular galaxies, and mergers, respectively. Previous studies also suggested that higher redshift samples experience faster quenching \citep[e.g.,][]{Goncalves2012}.

As mentioned in \Cref{Into}, previous studies found a significant fraction of X-ray detected AGN in the green valley region \citep[e.g.,][]{Cowie2008, Hickox2009, Schawinski2010, Povic2012, Wang2017, Gu2018, Lacerda2020} and suggested a link between AGN activity and the process of SF quenching, moving a galaxy from the blue cloud to the red sequence. AGN negative feedback has been proposed for faster quenching processes, being able to quench star formation on a scale of tens to hundreds of millions of years \citep[e.g.,][]{Barro2013, Dubois2013, Bongiorno2016, Smethurst2016, Kocevski2017}. 

In this work, for FIR emitters in the green valley, we obtain on average faster measured star-formation quenching time-scales of an order of $\sim$\,70\,Myr, for both AGN and non-AGN galaxies. Although 50\% of the AGN sample occupies a smaller range of SF time-scales (53\,Myr\,-\,103\,Myr) in comparison to non-AGN (36\,Myr\,-\,194\,Myr), on average we don't see significant differences between AGN and non-AGN. Therefore, star-formation time-scales measurements are also in line with other results shown in this paper, and in \cite{Mahoro2017, Mahoro2019}, where we do not see clear signs of  negative AGN feedback in our sample of X-ray detected FIR AGN.

\section{Summary}

In this paper, we have explored the properties of stellar populations and ages using a sub-sample of FIR AGN and non-AGN green valley galaxies selected in \cite{Mahoro2017, Mahoro2019} from the COSMOS survey at redshifts $\rm{0.6\,<\,z\,<\,1.0}$. We fit the spectral absorption features and continua of the green valley galaxies to study their stellar populations and ages using SSPs of \cite{Bruzual2003} utilising the STARLIGHT code and make use of spectral indicators, such as the 4000 $\rm{\AA}$ break and the H$\rm{\delta_{A}}$, as diagnostics of the past star formation histories of the galaxies. We also observed the location of the selected sub-sample on the main sequence of star formation in relation to their morphology and average stellar ages. Finally, we measured star-formation quenching time-scales of both AGN and non-AGN sub-samples. 

Our main results are summarised as follows:
\begin{enumerate}
\item Both FIR AGN and non-AGN green valley galaxies are dominated by intermediate and old stellar populations, where FIR AGN have a slightly higher fraction of intermediate stellar populations in comparison to non-AGN (67\% vs. 53\%, respectively) and a lower fraction of old stellar populations (23\% vs. 36\%, respectively). Only a small fraction of $\rm{\sim}$10\% of both AGN and non-AGN show young stellar populations.
\item Both AGN and non-AGN light-weighted stellar ages span a similar range, where the median age of AGN and non-AGN are $\rm{\log t\,=\,8.5\,[yr]}$ and $\rm{\log t\,=\,8.4\,[yr]}$, respectively. Both AGN and non-AGN  also have similar median mass-weighted stellar ages of $\rm{\log t\,=\,9.1\,[yr]}$.
\item We observed a similar trend in both samples where the star formation history of most of the sources (62\% of AGN and 66\% of non-AGN) appear to be explained by a mix of bursts and continuous star formation. On the other hand, a slightly higher fraction of AGN (38\%) compared to non-AGN (27\%) might have experienced a burst of star formation over the past 2 Gyr.
\item For non-AGN, we observed that galaxies with younger stellar ages are located mainly on the main sequence of star formation, while those with older ages are found below the MS, as expected. This trend is however not clear in the case of AGN, but the result could be affected by the small number of sources.
\item Finally, we found in both the AGN and non-AGN green valley galaxies similar star formation quenching time-scales of $\sim70$ Myr.
\end{enumerate} 

All of the results above are in line with those obtained in \cite{Mahoro2017, Mahoro2019}, where in the analysed sample of FIR AGN in the green valley we do not see clear signs of AGN negative feedback, as has been suggested in previous optical studies. 

\section*{Acknowledgements}
We thank Riffel Rogério for accepting to review this paper, and for giving us constructive and useful comments that improved the manuscript. This work is based on the research supported by the National  Research Foundation of South Africa (Grant Numbers 110816 and 132016). AM gratefully acknowledge financial support from the Swedish International Development Cooperation Agency (SIDA) through the International Science Programme (ISP) - Uppsala University to University of Rwanda through the Rwanda Astrophysics, Space and Climate Science Research Group (RASCSRG). MP acknowledges financial supports from the Ethiopian Space Science and Technology Institute (ESSTI) under the Ethiopian Ministry of Innovation and Technology (MoIT), from the Spanish Ministerio de Ciencia e Innovaci\'on - Agencia Estatal de Investigaci\'on through projects PID2019-106027GB-C41 and AYA2016-76682C3-1-P, and the State Agency for
Research of the Spanish MCIU through the Center of Excellence Severo Ochoa award to the Instituto de Astrof\'isica de Andaluc\'ia (SEV-2017-0709). PV acknowledges support from the  National  Research Foundation of South Africa. We are grateful to the COSMOS survey, LEGA-C survey  and all scientists whose data have been used in this research for making them publicly available and free. We are grateful to STARLIGHT team for making their code public. We are also grateful to the python, Virtual Observatory, and TOPCAT teams for making their packages freely available to the scientific communnity. 

\section*{Data Availability}
The data used in this work are available in the second public data release of LEGA-C \citep{Straatman2018}, available on the LEGA-C website.
\bibliographystyle{mnras}
\bibliography{Stellar_pop_paper} 
\end{document}